\documentclass[
superscriptaddress,
twocolumn,
amsmath,amssymb,
aps,
pra,
]{revtex4-1}

\usepackage{graphicx}

\usepackage{epsfig}
\usepackage{graphicx}
\usepackage{amsmath}
\usepackage{amsthm}
\usepackage{bm}
\usepackage{layout}
\usepackage{float}
\usepackage{dsfont}

\usepackage{amsfonts}

\begin{document}

\newcommand{\ket}[1]{\left | #1 \right\rangle}
\newcommand{\bra}[1]{\left \langle #1 \right |}
\newcommand{\half}{\mbox{$\textstyle \frac{1}{2}$}}
\newcommand{\smallfrac}[2][1]{\mbox{$\textstyle \frac{#1}{#2}$}}
\newcommand{\Tr}{\mathrm{Tr}}
\newcommand{\braket}[2]{\left\langle #1|#2\right\rangle}
\newcommand{\proj}[1]{\ket{#1}\bra{#1}}
\renewcommand{\epsilon}{\varepsilon}

\title{Paradoxical consequences of multipath coherence: \\ perfect interaction-free measurements}

\author{Z. Zhao}
\affiliation{School of Physical and Mathematical Sciences, Nanyang Technological University, Singapore 637371}

\author{S. Mondal}
\affiliation{School of Physical and Mathematical Sciences, Nanyang Technological University, Singapore 637371}
\affiliation{Indian Institute of Science, Bangalore, Karnataka 560012, India}

\author{M. Markiewicz}
\affiliation{Institute of Physics, Jagiellonian University, \L ojasiewicza 11, 30-348 Krak\'ow, Poland}

\author{A. Rutkowski}
\affiliation{Institute of Theoretical Physics and Astrophysics, Faculty of Mathematics, Physics and Informatics,
	University of Gda\'nsk, 80-308 Gda\'nsk, Poland}

\author{B. Daki\'c}
\affiliation{Institute of Quantum Optics and Quantum Information, Austrian Academy of Sciences, Boltzmanngasse 3, Vienna A-1090, Austria}
\affiliation{Vienna Center for Quantum Science and Technology (VCQ), Faculty of Physics, Boltzmanngasse 5, University of Vienna,Vienna A-1090, Austria}

\author{W. Laskowski}
\affiliation{Institute of Theoretical Physics and Astrophysics, Faculty of Mathematics, Physics and Informatics,
	University of Gda\'nsk, 80-308 Gda\'nsk, Poland}

\author{T. Paterek}
\affiliation{School of Physical and Mathematical Sciences, Nanyang Technological University, Singapore 637371}
\affiliation{MajuLab, CNRS-UCA-SU-NUS-NTU International Joint Research Unit, UMI 3654, Singapore 117543}

\date{\today}

\begin{abstract}
Quantum coherence can be used to infer the presence of a detector without triggering it. Here we point out that,
according to quantum mechanics, such interaction-free measurements cannot be perfect, i.e., in a single-shot
experiment one has strictly positive probability to activate the detector. We formalize the extent to which
such measurements are forbidden by deriving a trade-off relation between the probability of activation and the
probability of an inconclusive interaction-free measurement. Our description of interaction-free measurements
is theory independent and allows derivations of similar relations in models generalizing quantum mechanics.
We provide the trade-off for the density cube formalism, which extends the quantum model by permitting
coherence between more than two paths. The trade-off obtained hints at the possibility of perfect interaction-free
measurements and indeed we construct their explicit examples. Such measurements open up a paradoxical
possibility where we can learn by means of interference about the presence of an object in a given location without
ever detecting a probing particle in that location. We therefore propose that absence of perfect interaction-free
measurement is a natural postulate expected to hold in all physical theories. As shown, it holds in quantum
mechanics and excludes the models with multipath coherence.\\\\
DOI: 10.1103/PhysRevA.98.022108
\end{abstract}

\maketitle

A sample is seen under the microscope due to photons
scattered from it. Similarly, essentially all our knowledge about
the physical world comes from probes directly interacting with
the objects of interest. Yet, quantum mechanics offers another
possibility for enquiring whether an object is present at a given
location—the interaction-free measurement~\cite{Elitzur1993}.
It is possible
by interferometric techniques to prepare a single quantum particle
in superposition having one arm in a suspected location of
the object and with the measurement scheme which, from time
to time, identifies the presence of the object arguably without
directly interacting with it~\cite{Elitzur1993,QuantOpt.6.119,PhysRevLett.74.4763,HAFNER1997563,PhysRevA.57.3987,AppPhysB.123.12,PhysRevA.90.042109,Paraoanu2006}.
We ask here if interaction-free
measurements could be made perfect and provide nontrivial
information about the presence of the object, even if in each
and every run the particle and the object to be detected do not
interact directly. Within quantum formalism the answer is negative,
for which we provide an elementary argument as well as a
quantitative relation covering this conclusion as a special case.\\
\indent
One could therefore say that we have identified yet another
no-go theorem for quantum mechanics similar to no-cloning~\cite{no-cloning,PLA.92.271}, 
no-broadcasting~\cite{PhysRevLett.76.2818,PhysRevLett.100.090502} or no-deleting~\cite{no-deleting}.
Their
importance comes from pinpointing special features of the
quantum formalism (and the world) which can then be preserved
or relaxed one by one when studying candidate physical
theories. In this spirit, here we explore the possibility of perfect
interaction-free measurements in the framework of density
cubes~\cite{NJP.16.023028}.
The basic idea behind this framework is to represent
states by higher-rank tensors, density cubes, in direct analogy
to quantum mechanical density matrices. In this way, more
than two classically exclusive possibilities can be coherently
coupled, giving rise to genuine multipath interference absent in
quantum mechanics~\cite{NJP.16.023028,ModPhysLettA.9.3119}.
The particular interferometer employed to theoretically demonstrate the multipath interference
has a feature, also noted by Lee and Selby~\cite{FoundPhys.47.89}, 
that the particle
is never found in one of the paths inside the interferometer
but the presence of a detector in that path affects the final
interference fringes. It is exactly this property that we shall
exploit for the perfect interaction-free measurement.\\
\indent The observation that quantum mechanics does not give
rise to multipath coherence was made for the first time about
20 years ago~\cite{ModPhysLettA.9.3119}
and was linked to the validity of Born’s
rule: since the number of particles around a given point on
the screen is proportional to the square of the sum of the
probability amplitudes, only products of two amplitudes are
responsible for the interference. Experiments were set up to
look for genuine multipath interference and to test the Born
rule~\cite{Science.329.418,FoundPhys.42.742,NewJPhys.14.113025,NewJPhys.19.033017}.
In addition to being of fundamental interest, these
experiments also have practical implications, as it has been
shown that multipath interference provides an advantage over
quantum mechanics in the task called the ``three collision problem''~\cite{FoundPhys.47.89} and may be advantageous over quantum algorithms~\cite{NJP.18.033023}, although this is not the case in searching~\cite{NJP.18.093047}.
Up to now essentially all experimental data confirms the absence of genuine multipath interference and the consistency of the Born rule.These findings are also supported by additional theoretical
research. Namely, models with genuine multipath interference
were shown to be at variance with a number of postulates:
purity principles~\cite{NJP.16.123029,Entropy.19.253,1701.07449}, tomography via single-path and double-path experiments~\cite{FoundPhys.41.396,UdudecThesis}, possibility of defining composite
systems ~\cite{JPhys.41.355302} and experiments giving a definite (single) outcome ~\cite{1611.06461}.
The present contribution adds to this line of research. 
We identify paradoxical consequences of particular models with multi-path coherence that are phrased solely in operational terms and hence make the models highly unlikely to describe natural processes.

The paper is organized as follows. In Sec.~\ref{SEC_IFM} we introduce
interaction-free measurements and formally define perfect
interaction-free measurement in a theory-independent way.
We show that in all models where processes are assigned
probability amplitudes, satisfying natural composition laws,
there are no perfect interaction-free measurements and also
no genuine multipath interference. Furthermore, we derive
within quantum formalism a trade-off relation characterizing
interaction-free measurements, which explicitly shows the
impossibility of such perfect measurements. We then move
to the density cube model and for completeness gather in
Sec.~\ref{SEC_CUBES} all its elements necessary for our purposes. Similarly
to the quantum case, we derive the trade-off relation within
the density cube model, which now opens up the possibility
of perfect interaction-free measurement. Sec.~\ref{SEC_PIFM} provides
explicit examples of such measurements. The first example
uses a three-path interferometer having the property that the
particle is never found in the path where we place the detector,
but the interaction-free measurement fails $50\%$ of the time.
(We prove that this cannot be improved using the class of
interferometers considered.) In next example, we provide an
$N$-path interferometer giving rise to perfect interaction-free
measurement and a vanishing probability of failure in the limit
 $N \to \infty$. We conclude in Sec.~\ref{SEC_CONCLUDE}.


\section{Interaction-free measurements}
\label{SEC_IFM}

\begin{figure}[!t]\centering
\hspace{2cm}\includegraphics[width=9cm]{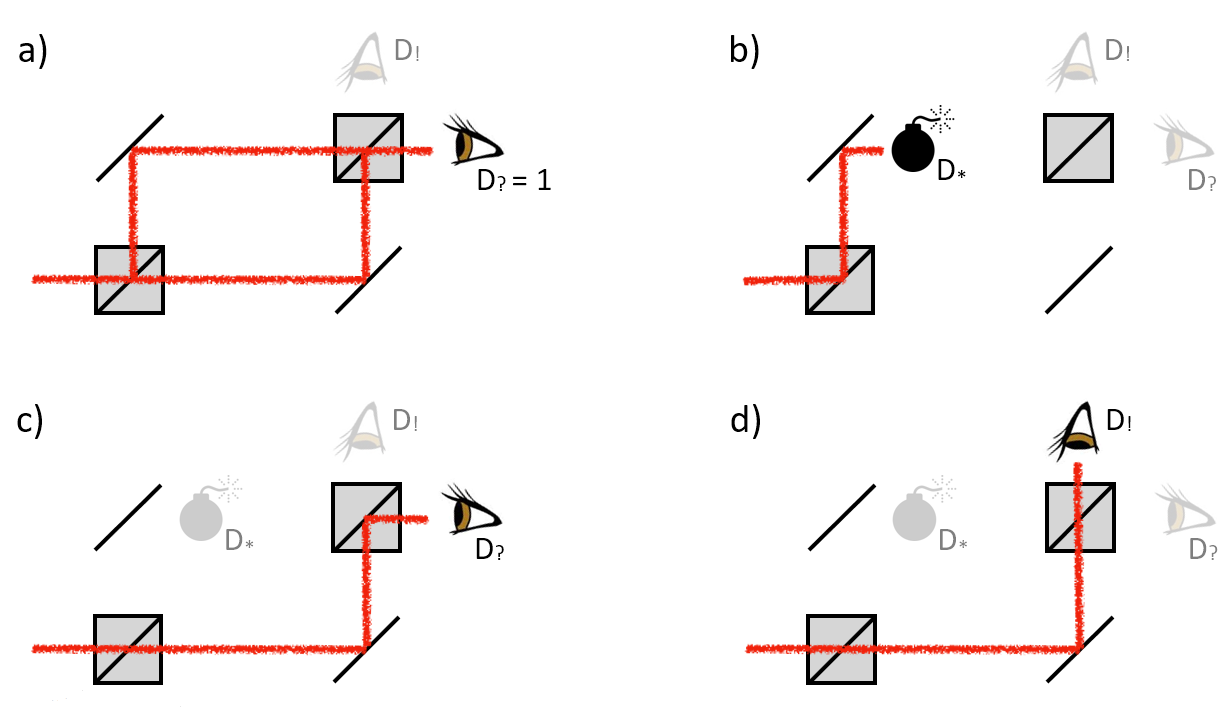}
\caption{Interaction-free measurement and the relevant parameters.
A tuned Mach-Zehnder interferometer, as in panel (a), is moved to
the location where there might be a bomb in the upper arm. The probe
particle triggers the bomb, single-particle-sensitive detector $D_*$, via
path (b) with probability $P_*$. The measurement is inconclusive if the
particle takes path (c) because the detector labeled $D_?$ also fires when
there is no bomb, see (a). The probability of an inconclusive result is
denoted by $P_?$. Finally, the measurement succeeds if the top detector
clicks and this happens with probability $P_!$, see (d). The measurement
is termed interaction-free, because had the particle interacted with the
bomb, it would trigger it, and it did not.}
\label{FIG_MZ}
\end{figure}

We begin with the original scheme by Elitzur and Vaidman~\cite{Elitzur1993}.
The idea is presented and described in Fig.~\ref{FIG_MZ}.
The problem is famously dramatized by considering the presence or absence of a single-particle-sensitive bomb, the tradition we shall also follow.

For a general interferometer (with many paths and arbitrary
transformations replacing the beam splitters in Fig.~\ref{FIG_MZ}) one always
starts by tuning it to destructive interference in at least one
of its output ports. In this way, if the click in one of these ports
is observed we conclude that the bomb is present in the setup.
This constitutes a successful interaction-free measurement and
we denote its probability by $P_!$. If the particle emerges in any
other output port, we cannot make any definite statement as
this happens both in the absence and presence of the bomb. The
result is therefore inconclusive and we denote its probability by
$P_?$. Finally, if the bomb is present, the probe particle triggers it
with probability $P_*$. Clearly we have exhausted all possibilities
and therefore
\begin{equation}
P_* + P_? + P_!  = 1.
\label{EQ_P_NORM}
\end{equation}

\subsection{Perfect interaction-free measurement}

We call an interaction-free measurement perfect when statistics of single-shot experiments with the same interferometer shows
\begin{equation}
P_* = 0 \quad \textrm{ and } \quad P_? < 1,
\label{PIFM}
\end{equation}
i.e., when the bomb never explodes, and yet from time to time,
we are certain it was there. The paper ends with an example
where both of these probabilities are zero.

It should be emphasized that this definition involves only
probabilities in certain experimental scenarios and hence it
is independent of the underlying physical theory. Now we
show that a broad class of physical models, including quantum
mechanics, does not permit perfect interaction-free measurements.


\subsection{No perfect interaction-free measurements in amplitude models}
In quantum mechanics, the condition $P_* = 0$ implies a
vanishing probability amplitude for the particle to propagate
along the path of the bomb. This means that the first (generalized)
beam splitter never sends the particle to that path
and hence it is irrelevant whether one places a bomb there
or not, i.e., $P_? = 1$. The same conclusion holds in any theory
that assigns probability amplitudes to physical processes and
demands that the vanishing probability of the process implies
a vanishing amplitude, e.g., the probability is an arbitrary
power of the amplitude. It is intriguing in the present context
that many of such models do not give rise to multipath
coherence. If the amplitudes are complex numbers (or even
pairs of real numbers), their natural composition laws lead to
Feynman rules, i.e. probability $\sim$ amplitude$^2$~\cite{IntJThPhys.27.543,PRA.81.022109}.
Sorkin's original argument then demonstrates the absence of multi-path interference~\cite{ModPhysLettA.9.3119}.

This elementary argument excludes the possibility of perfect
interaction-free measurements in quantum mechanics. We
therefore ask to what extent are such measurements forbidden.
The answer is phrased as a trade-off relation between the
probability of detonation and the probability of an inconclusive
result. It shows that the inconclusive result happens more and
more often with a decreasing probability of triggering the
bomb.


\subsection{Quantum trade-off}

\begin{figure}[!t]
\centering
\includegraphics[width=6cm]{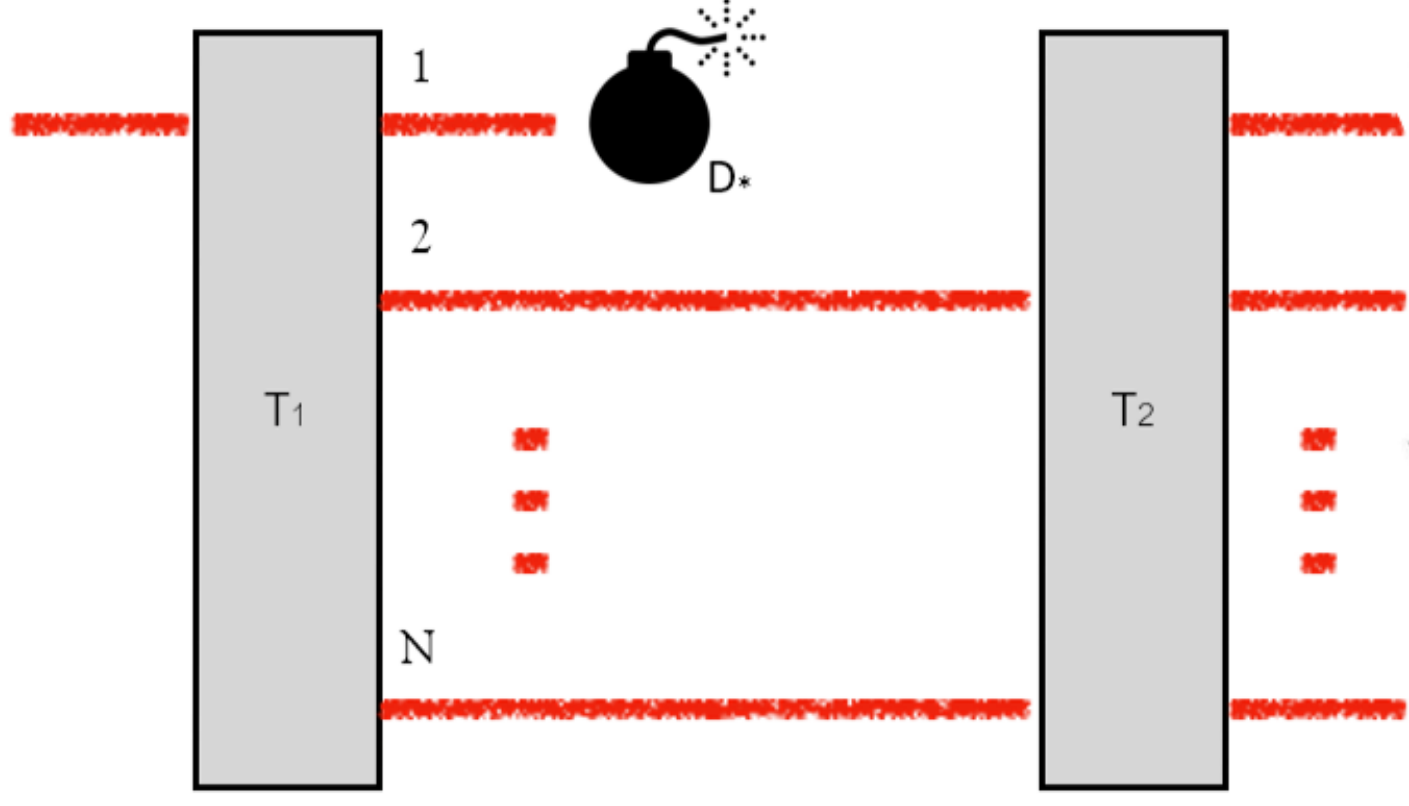}
\caption{A general interferometer used to derive the trade-off
	relations. The particle is injected into the first path from the left.
	It enters the interferometer via transformation $\mathcal{T}_1$ and leaves it via
	transformation $\mathcal{T}_2$. Inside the interferometer the bomb is present in
	the first path.}
\label{FIG_IFM}
\end{figure}

Consider a general interferometer as shown in Fig.~\ref{FIG_IFM}.
We present the trade-off between $P_*$ and $P_?$ for arbitrary mixed quantum states but keeping the second transformation unitary, i.e. $\mathcal{T}_2 = \mathcal{U}_2$.
Let us denote by $\rho$ the density matrix of the particle inside the interferometer, right after the first transformation.
The probability to trigger the bomb is given by
\begin{equation}
P_* = \bra{1} \rho \ket{1}.
\end{equation}
If the bomb is not triggered, the state $\rho$ gets updated to $\tilde \rho$ satisfying:
\begin{equation}
\tilde \rho = \frac{1}{1 - P_*} (\mathds{1} - \proj{1}) \rho (\mathds{1} - \proj{1}),
\label{EQ_TILDER}
\end{equation}
where $\mathds{1}$ is the identity operator in the space of density matrices.
Accordingly, the particle at the output of the interferometer is described by $\mathcal{U}_2 \,\tilde \rho \,\mathcal{U}_2^\dagger$.
The probability of an inconclusive result is given by the chance that now the particle is observed at those output ports $\ket{s}$ in which it might be present if there was no bomb:
\begin{equation}
P_? = (1- P_*) \sum_s \bra{s} \mathcal{U}_2 \,\tilde \rho\, \mathcal{U}_2^\dagger \ket{s},
\label{EQ_P?T}
\end{equation}
where we multiplied by $(1- P_*)$ to account for the renormalisation in $\tilde \rho$.
In Appendix~\ref{SEC_APP_QUANTUM} we derive the following trade-off relation:
\begin{eqnarray}
P_? & \ge & 1 - 2 P_* + P_* \bra{1} E(\rho) \ket{1} \nonumber \\
& \ge & (1 - P_*)^2, \label{EQ_QUANTUMT}
\end{eqnarray}
where $E(\rho)$ is the projector on the support of $\rho$, i.e. $E(\rho) = \sum_r \proj{r}$ for $\rho = \sum_r p_r \proj{r}$.
The last inequality in (\ref{EQ_QUANTUMT}) follows from convexity, $\bra{1} E(\rho) \ket{1} \ge \bra{1} \rho \ket{1}$. 
We now discuss special cases of this trade-off in order to illustrate the tightness of the bound and for future comparison with the model of density cubes.

First of all, due to convexity, the lower bound is saturated
by pure states. In other words, pure states are the best for
interaction-free measurements. Note also that in quantum
formalism, by starting with a pure state $\rho$ one always obtains
a pure state $\tilde \rho$ after the measurement. It turns out that the cube
model does not share this property.

Any density matrix $\rho$ that does not contain coherence to the state $\ket{1}$, i.e. has vanishing off-diagonal elements in the first row and column when $\rho$ is written in a basis including $\ket{1}$, is useless for interaction-free measurements.
If there is no coherence to state $\ket{1}$, then either (i) one of the eigenvectors of $\rho$ is this state or (ii) all the eigenvectors are orthogonal to $\ket{1}$.
In the case (i) we find $\bra{1} E(\rho) \ket{1} = 1$ and hence:
\begin{equation}
P_? \ge 1 - P_*.
\end{equation}
This combined with the normalisation condition (\ref{EQ_P_NORM}), means that there is no place for a successful interaction-free measurement, i.e. $P_! = 0$.
In the case (ii) we note that $P_* = 0$ and hence the lower bound in (\ref{EQ_QUANTUMT}) already shows that $P_? = 1$.
This demonstrates quantitatively the impossibility of perfect interaction-free measurements.
Finally, we note that the trade-off just derived holds for an
arbitrary interferometer (with the second transformation being
unitary) and that it is independent of the number of paths. For
example, the inequality (\ref{EQ_QUANTUMT}) is saturated by taking the discrete
Fourier transform as both transformations in the interferometer
with an arbitrary number of paths. This again will differ in the
density cube model.


\section{Density cubes}
\label{SEC_CUBES}
The trade-off relation just derived captures the impossibility
of perfect interaction-free measurement in the quantum formalism.
We show here that their absence is a natural postulate
which disqualifies certain extensions of quantum mechanics,
namely, the density cube model~\cite{NJP.16.023028}. This model has been
introduced in order to incorporate the possibility of multipath
coherence, and we shall first say a few words about where
exactly could such an extension show up in an experiment.
Sorkin introduced the following classification~\cite{ModPhysLettA.9.3119}. Quantum
mechanics gives rise to second-order interference because the
interference pattern observed on the screen behind two open
slits, I12, cannot be understood as a simple sum of patterns
when each individual slit is closed, i.e.,  $I_{12} - I_1 - I_2 \ne 0$.
However, the interference fringes observed in a triple-slit
experiment are always reducible to a simple combination
of double-slit and single-slit patterns, namely, $I_{123} = I_{12} + I_{13} + I_{23} - I_1 - I_2 - I_3$. Similar statements hold for higher numbers of slits. Why does quantum mechanics “stop” at
second-order interference? How would a model that gives rise
to third-order and higher-order interference look? The density
cube formalism provides the answer to the latter question. In
principle it could show up in triple-slit experiments. However,
the present paper finds that this is unlikely because the cubes
allow for perfect interaction-free measurements.

In order to keep the present work self-contained, we first review the
elements of the cubes model.We then derive the trade-off
relation between $P_?$ and $P_*$ within the density cubes framework, which hints at the possibility of perfect interaction-free measurement. Finally, we provide explicit examples of such
measurements.


\subsection{Probability}
The main difference between quantum mechanics and the
cubes framework is that instead of a density matrix, one assigns
a rank-$3$ tensor (density cube) to a given physical configuration.
The density cube $C$ can have complex elements $C_{jkl} \in \mathbb{C}$.
The density cubes are assumed to be Hermitian in the sense
that exchanging two indices produces a complex conjugated
element, e.g.,
\begin{equation}
C_{jkl} = C_{kjl}^*.
\end{equation}
Hermitian cubes form a real vector space with inner product
\begin{equation}
(M,C) = \sum_{j,k,l = 1}^N M_{jkl}^* C_{jkl},
\label{EQ_BORN}
\end{equation}
where each index of the tensor runs through values $1,\dots,N$.
Therefore, one naturally defines the probability of observing
an outcome corresponding to cube $M$ in a measurement on
a physical object described by cube $C$ by the above inner
product. This is in close analogy to the Born rule in quantum
mechanics, which in the same situation assigns probability
$\Tr(M C) = \sum_{j,k = 1}^N M_{jk}^* C_{jk}$, with $M$ and $C$ being density
matrices. In this way the model of the density cubes extends
the self-duality between states and measurements present in
quantum mechanics~\cite{NJP.12.033034,1110.6607}.


\subsection{States}
We shall consider two types of density cubes: the quantum
cubes which represent quantum states in the density cube
model and nonquantum cubes (with triple-path coherence)
that extend the quantum set. The former are constructed from
quantum states and are in one-to-one relationwith the quantum
states. While nonquantum cubes are also constructed starting
from a quantum state, one can choose various combinations
for the triple-path coherence terms to construct several distinct
nonquantum cubes corresponding to a given quantum state.

The sets of allowed density cubes and their transformations
are not yet fully characterized and it is not our aim
to characterize them in this paper. We will rather focus on
specific density cubes and transformations, which will be
shown to be consistent and will produce perfect interaction-free
measurements.

\subsubsection{Quantum cubes}

Consider the following mapping between a density matrix $\rho$ and a cube $C^Q$:
\begin{equation}
\begin{array}{rclcl}
	C_{jjj}^Q & = & \rho_{jj}, & & \\
	C_{jjk}^Q & = & \sqrt{\frac{2}{3}} \, \textrm{Re}(\rho_{jk}), & \textrm{ for } & j < k, \\
	C_{jkk}^Q & = & \sqrt{\frac{2}{3}} \, \textrm{Im}(\rho_{jk}),& \textrm{ for } & j<k, \label{EQ_Q_CUBE} \\
	C_{jkl}^Q & = & 0, & \textrm{ for } & j \ne k \ne l.
\end{array}
\end{equation}
Note that all the terms $C_{jkl}^Q$ where the three indices are different
are set to zero, meaning that these cubes do not admit any three-path
coherence. The remaining elements can be computed
using the Hermiticity rule. This mapping preserves the inner
product between the states, and hence quantum mechanics and
the density cube model with this set of cubes are physically
equivalent.

\subsubsection{Non-quantum cubes}

We now extend the set of quantum cubes and allow for non-trivial triple-path coherence by mapping every quantum state $\rho$ to the following family of cubes:
\begin{eqnarray}
C_{jjj} & = & \frac{1}{N-1}(1-\rho_{jj}),\nonumber \\
C_{jjk} & = & \sqrt{\frac{2}{3}} \frac{1}{N-1} \, \textrm{Re}(\rho_{jk}),\quad  \textrm{ for } \quad j < k, \nonumber\\
C_{jkk} & = & \sqrt{\frac{2}{3}} \frac{1}{N-1} \, \textrm{Im}(\rho_{jk}),\quad  \textrm{ for } \quad j < k, \nonumber\\
C_{1jk}(\gamma) & = & \sqrt{\frac{1}{3}} \frac{1}{N-1} \, \omega^{f(\gamma,j,k)}, \quad  \textrm{ for } \quad 1<j<k, \nonumber\\\label{eq:rho1jk}
\end{eqnarray} 
where $\omega = \exp(- i 2\pi / N)$ is the $N$th complex root of unity and $f(\gamma,j,k) = \{ 1,\dots, N \}$.
The parameter $\gamma = 1, \dots, N$ enumerates different cubes that can be constructed from a given quantum state.
Again, the remaining elements can be completed using the Hermiticity rule.
We provide explicit examples of interesting non-quantum cubes in Sec.~\ref{SEC_PIFM} and Appendix~\ref{SEC_APP_ONEQ}.
Note that for simplicity we choose to place the bomb in the first
path of the interferometer and therefore consider cubes where
the three-path coherence involves only the first path (labeled by index $1$) and two other paths. All the terms $C_{jkl}$, with three different indices, each of which is strictly greater than $1$, are
set to zero.


\subsection{Measurement}
We shall only be interested in enquiring about the particle’s
path at various stages of the evolution. Furthermore, we will
focus on checking whether the particle is in the first path or not.
Clearly this measurement is allowed in quantum mechanics,
and we choose vector $(1 \, 0 \, 0 )^T$ to represent the particle moving
along the first (out of three) paths inside the interferometer. The
corresponding quantum cube looks as follows, see Eq.~(\ref{EQ_Q_CUBE}), in
the case of the triple-path experiment:
\begin{eqnarray}
M_1 & = & \left\{ 
\left( \begin{array}{ccc} 1 & 0 & 0\\
0 & 0 & 0 \\
0 & 0 & 0
\end{array} \right),
\left( \begin{array}{ccc} 0 & 0 & 0 \\
0 & 0 & 0 \\
0 & 0 & 0
\end{array} \right),
\left( \begin{array}{ccc} 0 & 0 & 0 \\
0& 0 & 0 \\
0 & 0 & 0
\end{array} \right)
\right\},\nonumber\\ 
\end{eqnarray}
where the three $3 \times 3$ matrices describing the cube have elements $C_{1jk}$, $C_{2jk}$ and $C_{3jk}$, respectively.
The probability that a particle described by cube $C$ is found in the first path is $(M_1,C)$.

It is essential to the interaction-free measurement to describe
the state of the particle after it has \emph{not} been found in
a particular path. Here the model of density cubes follows
quantum mechanics and it is assumed that the cube describing
the system changes as a result of measurement. If the particle is
found in the $n$th path, its state gets updated $C \to M_n$, where $M_n$ is the quantum cube corresponding to a particle propagating
along the $n$th path. If the particle is not found in the $n$th path,
the model follows the generalized L\"uder's state update rule:
it erases from the cube all elements $C_{jkl}$ with $j,k,l = n$, and
renormalizes the remaining elements. Following our three-path
example, if the particle is not found in the first path its generic
cube $C$ gets updated to cube $\tilde C$ with elements
\begin{eqnarray}
\left\{ 
\left( \begin{array}{ccc} 0 & 0 & 0\\
0 & 0 & 0 \\
0 & 0 & 0
\end{array} \right),
\left( \begin{array}{ccc} 0 & 0 & 0 \\
0 & \tilde C_{222} & \tilde C_{223} \\
0 & \tilde C_{232} & \tilde C_{233}
\end{array} \right),
\left( \begin{array}{ccc} 0 & 0 & 0 \\
0& \tilde C_{322} & \tilde C_{323} \\
0 & \tilde C_{332} & \tilde C_{333}
\end{array} \right)
\right\},\nonumber\\
\label{EQ_RED}
\end{eqnarray}
where
\begin{equation}
\tilde{C}_{jkl} = \frac{1}{1- C_{111}} C_{jkl},
\label{rhotildaelements}
\end{equation}
is the cube element renormalised by the probability that the particle is not in the first path.

At this stage we must ensure that all postmeasurement cubes
are allowed within the model. This is immediately clear if one
begins with a quantum cube. For the nonquantum cubes we
note that we only consider those cubes which have three-path
coherence to the first path and we only enquire whether the
particle is in the first path or not. If the measurement does
not find the particle in the first path, all these coherences are
updated to zero and accordingly, the postmeasurement cube is
a quantum one.


\subsection{Cubes trade-off}

\begin{figure}[!t]\centering
\includegraphics[width=9cm]{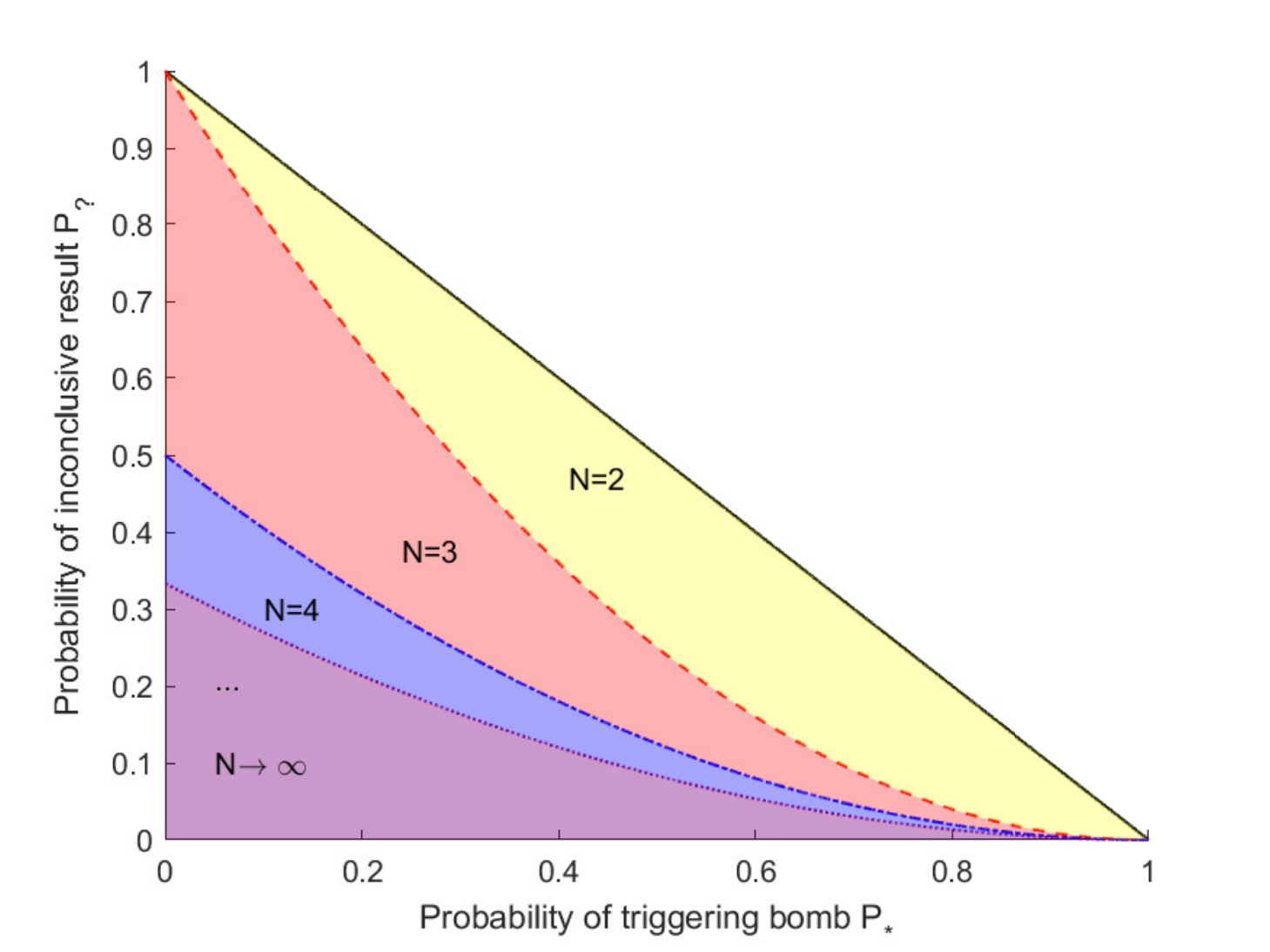}
\caption{Trade-off between the probability to trigger the bomb $P_*$
	and the probability of an inconclusive result $P_?$ within the cubes model
	and quantum mechanics. The straight line illustrates the trivial bound
 $P_* + P_? = 1$. All other region borders give lower bounds on the value
	of $P_?$ as a function of $P_*$. The available region for any $N$ (number of
	paths inside the interferometer) contains also the regions for all lower
	values of $N$. The quantum trade-off coincides with the case of $N = 2$.
	Perfect interaction-free measurements occur if the allowed values on
	the vertical axis are less than $1$.}
\label{FIG_REGIONS}
\end{figure}

We are now ready to present the trade-off relation between $P_*$ and $P_?$ for a general interferometer in Fig.~\ref{FIG_IFM}.
Our trade-off relation holds for the transformation $\mathcal{T}_2$ that preserves the inner product, 
while having an additional assumption on the structure of the cube $C$ describing the particle inside the interferometer right after $\mathcal{T}_1$.
We assume that after the particle has propagated through the whole interferometer in the case of no bomb, the cube at the output does not have any two-path and three-path coherence:
\begin{equation}
\mathcal{T}_2(C) = \sum_s p_s M_s.
\label{EQ_CUBESASSUMPTION}
\end{equation}
We ensure this is always fulfilled in our examples.
Similarly to the quantum case, the probabilities entering the trade-off are defined as follows:
\begin{eqnarray}
P_* & = & (M_1,C), \nonumber \\
P_? & = & (1- P_*) \sum_s (M_s,\mathcal{T}_2(\tilde C)),
\end{eqnarray}
where the cube $\tilde C$ represents the particle inside the interferometer after the measurement in the first path has not found the particle there [see Eq.~(\ref{EQ_RED})].
In Appendix~\ref{SEC_APP_CUBES} we derive the trade-off relation within the cubes model:
\begin{equation}
P_? \ge \frac{(1-P_*)^2}{N-1}.
\label{EQ_CUBET}
\end{equation}
It is illustrated in Fig.~\ref{FIG_REGIONS}.
One recognises that for $N=2$ this relation reduces to the one derived in quantum mechanics.
For two-path interferometers this is not surprising as in this case the density cube model reduces to standard quantum formalism~\cite{NJP.16.023028}.
For a higher number of paths, this relation emphasizes
that independence of the number of paths is a special quantum feature.

Relation (\ref{EQ_CUBET}) opens up the possibility of perfect interaction-free
measurements. Indeed, for all $N \ge 3$ one finds that the
right-hand side is strictly less than 1 even if $P_* = 0$. Furthermore,
both probabilities $P_*$ and $P_?$ can in principle be brought
to zero in the limit $N \to \infty$. In the next section we provide
explicit examples of perfect interaction-free measurements
which achieve the lower bound set by the trade-off relation (\ref{EQ_CUBET}).


\subsection{Examples of perfect interaction-free measurements}
\label{SEC_PIFM}
We present in detail the workings of the perfect interaction-free
measurement in the case of a three-path interferometer
with emphasis on the features departing from the quantum
formalism. The subsequent section provides the generalization
to $N$ paths. We discuss the main idea here and refer to Appendix~\ref{SEC_APP_T} for the details.

\subsubsection{Three paths}
\begin{figure}[!b]\centering
	\includegraphics[width=8cm]{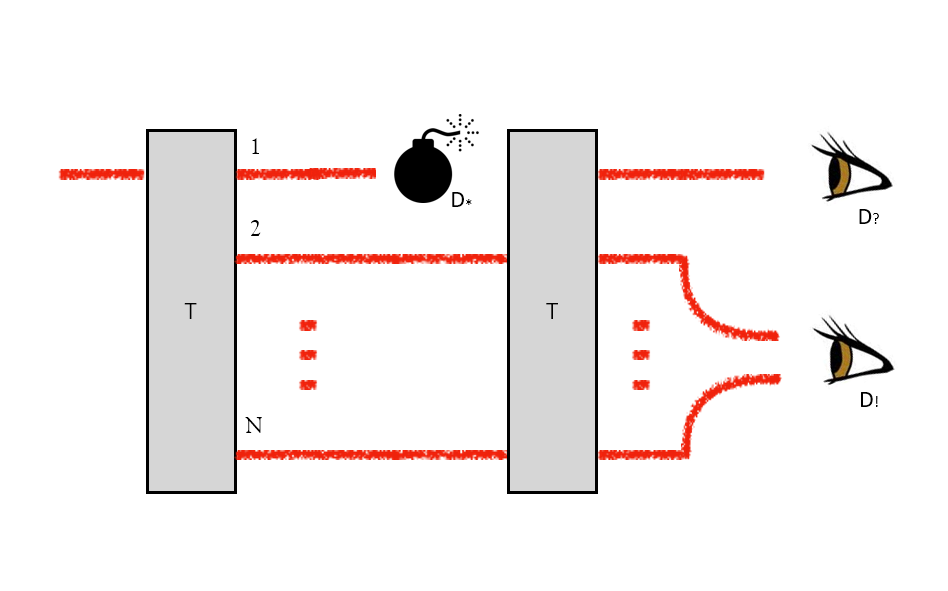}
	\caption{A perfect interaction-free measurement within the density
		cube model. Both transformations are the same and they have the
		property that $\mathcal{T}^2 = \mathds{1}$. Therefore, if there is no bomb the particle which
		enters through the first path always leaves the interferometer along the
		first path. The cube describing the particle inside the interferometer
		has only triple-path coherences to the first path and yet vanishing
		element $C_{111}$. Therefore the particle is never found along the first
		path inside the interferometer and the bomb never detonates, $P_* = 0$.
		However, the presence of the bomb removes the triple-path coherences
		from the cube. In this case, the second transformation evolves the
		particle to the first output port only with probability $P_? = \frac{1}{2}$. See
		main text for details.}
	\label{FIG_PIFM3}
\end{figure}

Consider the setup described in Fig.~\ref{FIG_PIFM3}. 
The transformation $\mathcal{T}_1 = \mathcal{T}_2 = \mathcal{T} \circ \mathcal{D}$ is chosen to consist of (quantum mechanical) complete dephasing of two-path coherences, $\mathcal{D}$,
followed by the transformation $\mathcal{T}$ defined in Eq. (16) of Ref.~\cite{NJP.16.023028}, which we will here review for completeness. The reason behind this composition of operations is that $\mathcal{T}$ is only
defined on a subset of cubes and it might be that it is impossible to extend it consistently to the whole set of cubes. The role of the dephasing is then to bring an arbitrary cube to the subset on
which $\mathcal{T}$ is known to act consistently. The dephasing operation
is defined to remove completely all two-path coherences in
a cube and leave unaffected the diagonal elements $C_{nnn}$ and
triple-path coherences $C_{jkl}$ with all indices different. Since this
operation acts only on the quantum part of the cube it produces
allowed cubes. Transformation $\mathcal{T}$ has matrix representation
\begin{equation}
\mathcal{T} = \frac{1}{2} \left(
\begin{array}{ccccc}
0 & 1 & 1 & 1 & 1 \\
1 & 0 & 1 & \omega^* & \omega \\
1 & 1 & 0 & \omega & \omega^* \\
1 & \omega & \omega^* &1 & 0 \\
1 & \omega^* & \omega & 0 & 1
\end{array}
\right),
\end{equation}
when written in the following sub-basis of Hermitian cubes:
\begin{eqnarray}
B_1 & = & M_1, \qquad B_2 = M_2, \qquad B_3 = M_3, \nonumber \\
B_4 & = & \frac{1}{\sqrt{3}}
\left\{ 
\left( \begin{array}{ccc} 0 & 0 & 0\\
0 & 0 & 1 \\
0 & 0 & 0
\end{array} \right),
\left(\begin{array}{ccc} 0 & 0 & 0 \\
0 & 0 & 0 \\
1 & 0 & 0
\end{array} \right),
\left(\begin{array}{ccc} 0 & 1 & 0 \\
0 & 0 & 0 \\
0 & 0 & 0
\end{array} \right)
\right\}, \nonumber \\
B_5 & = & \frac{1}{\sqrt{3}}
\left\{ 
\left( \begin{array}{ccc} 0 & 0 & 0\\
0 & 0 & 0 \\
0 & 1 & 0
\end{array} \right),
\left(\begin{array}{ccc} 0 & 0 & 1 \\
0 & 0 & 0 \\
0 & 0 & 0
\end{array} \right),
\left(\begin{array}{ccc} 0 & 0 & 0 \\
1 & 0 & 0 \\
0 & 0 & 0
\end{array} \right)
\right\}.\nonumber\\\label{EQ_subbasis}
\end{eqnarray}
That is, given an arbitrary cube $C$ in this subspace, the transformation $\mathcal{T}$  acts upon it via ordinary matrix multiplication on the vector representation of $C$, i.e., a five-dimensional column
vector with $j$th component given by $(C,B_j)$. As already alluded
to, this subspace consists of cubes which have no two-path
coherences but solely three-path coherences and diagonal
terms. It is now straightforward to verify that $\mathcal{T}$ is an involution
in the considered subspace, i.e. $\mathcal{T}^2 = \mathds{1}$. Accordingly, if the
particle enters the interferometer through the first path, it is
always found in the first output port of the setup. This adheres
to our assumption (\ref{EQ_CUBESASSUMPTION}) as the output cube is simply $M_1$.
The transformation T is different from an arbitrary unitary
transformation as it produces triple-path coherence inside
the interferometer. The particle injected into the first path is
described by the cube $M_1$, which in the considered subspace
corresponds to the vector $(1 \, 0 \, 0 \, 0 \, 0)^T$, and one can verify that
the corresponding cube after application of $\mathcal{T}$ is $\mathcal{T}(M_1) = \frac{1}{2}(B_2 + B_3 + B_4 + B_5) = C $,
which is also given by
\begin{equation}
\frac{1}{2}\left\{ 
\left( \begin{array}{ccc} 0 & 0 & 0\\
0 & 0 & \frac{1}{ \sqrt{3}} \\
0 & \frac{1}{ \sqrt{3}} & 0
\end{array} \right),
\left(\begin{array}{ccc} 0 & 0 & \frac{1}{ \sqrt{3}} \\
0 & 1 & 0 \\
\frac{1}{ \sqrt{3}} & 0 & 0
\end{array} \right),
\left(\begin{array}{ccc} 0 & \frac{1}{ \sqrt{3}} & 0 \\
\frac{1}{ \sqrt{3}} & 0 & 0 \\
0 & 0 & 1
\end{array} \right)
\right\}.
\end{equation}
Note that it is a pure cube, i.e., $(C,C) = 1$, and it contains
solely three-path coherences and elements $C_{222}$ and $C_{333}$. The
essential feature we are utilizing for perfect interaction-free
measurement is the presence of these coherences even though
the probability to find the particle in the first path vanishes:
\begin{equation}
P_* = (M_1,C) = 0.
\end{equation}
A similar statement for quantum states does not hold. If
the probability to locate a quantum particle in the first path
vanishes, all coherences to this path must vanish, as otherwise
the corresponding density matrix has negative eigenvalues.

If the bomb is present inside the interferometer but is not
triggered, the state update rule dictates erasure of all elements $C_{jkl}$ with any of $j,k,l = 1$.
We obtain the following cube:
\begin{equation}
\tilde C = \left\{ 
\left( \begin{array}{ccc} 0 & 0 & 0\\
0 & 0 & 0 \\
0 & 0 & 0
\end{array} \right),
\left(\begin{array}{ccc} 0 & 0 & 0 \\
0 & \frac{1}{2} & 0 \\
0 & 0 & 0
\end{array} \right),
\left(\begin{array}{ccc} 0 & 0 & 0 \\
0 & 0 & 0 \\
0 & 0 & \frac{1}{2}
\end{array} \right)
\right\}.
\end{equation}
It contains no coherences whatsoever and it is mixed, i.e., $(\tilde C, \tilde C) = \frac{1}{2}$.
We started with a pure cube and post-selected a mixed one.
This is also not allowed within the quantum formalism, where any pure state $\ket{\psi} = \sum_{n = 1}^N \alpha_n \ket{n}$ gets updated to another pure state $|\tilde \psi \rangle = \sum_{n=2}^N \tilde \alpha_n \ket{n}$, with $\tilde \alpha_n = \alpha_n / \sqrt{1-|\alpha_1|^2}$.

Finally, we evolve $\tilde C = \frac{1}{2} M_2 + \frac{1}{2} M_3$ through the second transformation and find that $\mathcal{T}(\tilde C)$ is given by (dephasing has no effect here):
\begin{eqnarray*}
\Big\{ 
\left( \begin{array}{ccc} \frac{1}{2} & 0 & 0\\
0 & 0 & - \frac{1}{4\sqrt{3}} \\
0 & - \frac{1}{4\sqrt{3}} & 0
\end{array} \right) 
&,&
\left(\begin{array}{ccc} 0 & 0 & - \frac{1}{4\sqrt{3}} \\
0 & \frac{1}{4} & 0 \\
- \frac{1}{ 4 \sqrt{3}} & 0 & 0
\end{array} \right),\\
& &
\left(\begin{array}{ccc} 0 & - \frac{1}{ 4\sqrt{3}} & 0 \\
- \frac{1}{ 4\sqrt{3}} & 0 & 0 \\
0 & 0 & \frac{1}{4}
\end{array} \right)
\Big\}.
\end{eqnarray*}
The probability of an inconclusive result is given by the
probability that the particle is found in the first path, as it was
always there in the absence of the bomb, and therefore we find
\begin{equation}
P_? = (M_1, \mathcal{T}(\tilde C)) = \frac{1}{2}.
\end{equation}
This probability saturates the lower bound derived in Eq.~(\ref{EQ_CUBET}) for $N=3$ and hence the setup discussed is optimal.

\subsubsection{More than three paths}
We now generalize the above scheme to more than three
paths and show that the density cube model allows for perfect
interaction-free measurement, which in every run provides
complete information about the presence of the bomb. This
holds in the limit $N \to \infty$.
We shall now construct a set of $N$ pure orthonormal cubes
 $C^{(n)}$, which will then be used to provide the transformation
$\mathcal{T}$ of the optimal interferometer, that gives rise to the minimal
probability of an inconclusive result while keeping $P_* = 0$. We
set the modulus of all the three-path coherences within each
cube $C^{(n)}$ to be the same and choose its non-zero elements as
follows:
\begin{eqnarray}
C_{jjj}^{(n)} & = & \frac{1}{N-1}, \quad \textrm{ for } j \ne n, \nonumber \\
C_{1jk}^{(n)}  & = & \sqrt{\frac{1}{3}} \frac{1}{N-1} \, x_{jk}^{(n)}, \quad \textrm{ for } 1< j < k.
\end{eqnarray}
The other non-zero three-path coherences can be found from the Hermiticity rule.
In this way cube $C^{(n)}$ is represented by a set of phases $x_{jk}^{(n)}$.
We arrange the independent phases, i.e. the ones having $j < k$, into a vector $\vec x_n$.
The orthonormality conditions between the cubes are now expressed in the following equations
\begin{align}
(C^{(n)}, C^{(n)}) = 1  \iff  &| (\vec x_{n})_j | =1 \nonumber\\
&\textrm{ for all } n,j, \nonumber \\
(C^{(m)}, C^{(n)}) = 0  \iff  &(\vec{x}_{m},\vec{x}_{n})+(\vec{x}_{n},\vec{x}_{m})=2-N, \nonumber\\ 
&\textrm{ for all } m \ne n. 
\label{EQ_ONEQ}
\end{align}
where $(\vec x_{n})_j$ is the $j$th component of the vector $\vec x_n$.
Equations (\ref{EQ_ONEQ}) are solved in Appendix~\ref{SEC_APP_ONEQ}.
Let us write the solution in form of a matrix
\begin{equation}
X = (\vec x_1 \dots \vec x_N),
\end{equation}
having vectors $\vec x_n$ as columns.
We now show how to use it to construct the ``cube multiport'' transformation $\mathcal{T}$.

We assume the two transformations in the setup are the same and that $\mathcal{T}$ is defined solely on the subspace of Hermitian cubes which do not have any two-path coherences. 
The cubes forming the basis set for this subspace are as follows:
\begin{align}
	B_{jkl}^{(n)}  =  &\delta_{jn} \delta_{kn} \delta_{ln},\quad \textrm{ for } n = 1, \dots, N \label{EQ_SUBB} \\
	B^{(vw)}_{jkl}  = &\frac{1}{\sqrt{3}} \left( \delta_{j1}\delta_{kv}\delta_{lw}+\delta_{jw} \delta_{k1}\delta_{lv}+\delta_{jv}\delta_{kw}\delta_{l1} \right), \\&\textrm{ for } 1<v<w\leq N, \nonumber \\	
	B^{(wv)}_{jkl}  =&\frac{1}{\sqrt{3}} \left(\delta_{j1}\delta_{kw}\delta_{lv}+\delta_{jv} \delta_{k1}\delta_{lw}+\delta_{jw}\delta_{kv}\delta_{l1} \right), \\ &\textrm{ for } 1<v<w\leq N. \nonumber
\end{align}
One recognizes that the cubes in the first line are just the $M_n$
cubes describing the particle propagating along the $n$th path.
The cubes in the second line describe independent three-path
coherences, and the cubes in the third line their complex
conjugates. Altogether there are $d = N + (N-1)(N-2)$
cubes in this sub-basis, and hence the transformation $\mathcal{T}$ is
represented by a $d \times d$ matrix, which we then divide into blocks:
\begin{equation}
\mathcal{T} =
\left(
\begin{array}{c|c}
A & C \\
\hline
B & D
\end{array}
\right),
\label{eq:T}
\end{equation}
$A$ being a square $N \times N$ matrix, $D$ being a square matrix with dimension $(N-1)(N-2) \times (N-1)(N-2)$, and $B$ and $C$ being rectangular.
By imposing $\mathcal{T}(M_n) = C^{(n)}$, matrices $A$ and $B$ are fixed to
\begin{equation}
A = \frac{1}{N-1}
\left(
\begin{array}{cccc}
0&1&\cdots&1\\
1&0&\cdots&1\\
\vdots&\vdots&\ddots&\vdots\\
1&1&\cdots&0\\
\end{array} \right),
\qquad
B = \left(
\begin{array}{c}
X \\
X^*
\end{array}
\right).
\end{equation}
By further requiring involution $\mathcal{T}^2 = \mathds{1}$ and Hermiticity $\mathcal{T} = \mathcal{T}^\dagger$ one finds that
\begin{equation}
C = B^\dagger, \qquad D = \sqrt{\mathds{1} - BB^\dagger}.
\end{equation}
We show in Appendix~\ref{SEC_APP_T} that $\mathds{1} - BB^\dagger$ is a positive matrix,
which concludes our construction of $\mathcal{T}$. It turns out that
this is not the only way to construct the cube multiport
transformation and Appendix D provides other examples. All
of them transform the quantum cubes $M_n$  to the nonquantum
cubes $C^{(n)}$. Note that in the considered subspace $M_n$ are the
only pure quantum cubes, and one verifies that $C^{(n)}$ are the
only pure nonquantum cubes allowed. In this way $\mathcal{T}$ is shown
to act consistently, i.e., map cubes allowed within the model
to other allowed cubes.

\section{Conclusions}
\label{SEC_CONCLUDE}
We proposed a theory-independent definition of perfect
interaction-free measurement. It turns out that quantum mechanics
does not allow this possibility, which we show by an
elementary argument and by a quantitative relation. However,
it can be realized within the framework of density cubes~\cite{NJP.16.023028}.
This framework allows transformations that prepare triple-path
coherence involving a path where the probability of detecting
the particle is strictly zero. Nevertheless, this coherence can
be destroyed if the bomb (detector) is in the setup, leading to
a distinguishable outcome in a suitable one-shot interference
experiment. We emphasize that, in this paper, we study single-shot
experiments in contrast to the quantum Zeno effect where
the interferometer is used multiple times~\cite{PhysRevLett.74.4763,PhysRevA.90.042109}.

We postulate that perfect interaction-free measurements
should not be present in a physical theory as they effectively
allow deduction of the presence of an object in a particular
location without ever detecting a particle in that location. One
might also try to identify more elementary principles which imply
the impossibility of perfect interaction-free measurements.

In this context, we note that perfect interaction-free measurements
are consistent with the no-signaling principle (no
superluminar communication). In the density cube model it
is the triple-path coherence that is being destroyed by the
presence of the detector inside the interferometer. The statistics
of any observable measured on the remaining paths is the same,
independently of whether the detector is in the setup or not.
Hence the information about its presence can only be acquired
after recombining the paths together, which can be done at
most at the speed of light. The situation resembles that of
the stronger-than-quantum correlations satisfying the principle
of no-signaling~\cite{PR}. They are considered “too strong,” as
they trivialize communication complexity~\cite{vanDam,PhysRevLett.96.250401} or random access coding~\cite{ic}, and they are at variance with many
natural postulates~\cite{ic,ML,NJP.14.063024}. Similarly, we consider identifying
the presence of a detector without ever triggering it, i.e., a
perfect interaction-free measurement, as too powerful to be
realized in nature. Exactly which physical principles forbid
such measurements is, of course, an interesting question.

Finally, we wish to comment briefly on experimental tests
of genuine multipath interference. They are often described as
simultaneously testing the validity of Born's rule. Indeed, as
we pointed out here, this is the case for a broad class of models
which assign probability amplitudes to physical processes and
these amplitudes satisfy natural composition laws~\cite{IntJThPhys.27.543,PRA.81.022109}.
Other models, however, are possible, as exemplified by the
density cube framework. Within this framework the probability
rule is essentially the same as the Born rule in quantum
mechanics. [For its version for mixed states, see Eq.~(\ref{EQ_BORN}).]
Therefore, in general, tests of multipath interference should
be distinguished from validity tests of Born's rule.
\section*{Acknowledgement}
We thank Pawe{\l} B{\l}asiak, Ray Ganardi, Pawe{\l} Kurzy\'nski and Marek Ku\'s for discussions and the NTU-India Connect
Programme for supporting the visit of S.M. to Singapore. This
research is funded by the Singapore Ministry of Education
Academic Research Fund Tier 2, Project No. MOE2015-T2-2-034, and Narodowe Centrum Nauki (Poland) Grant
No. 2014/14/M/ST2/00818. W.L. is supported by Narodowe
Centrum Nauki (Poland) Grant No. 2015/19/B/ST2/01999.
M.M. acknowledges the Narodowe Centrum Nauki (Poland),
through Grant No. 2015/16/S/ST2/00447, within the FUGA 4
project for postdoctoral training.

\appendix

\section{Proof of the quantum trade-off}
\label{SEC_APP_QUANTUM}

Let us denote the eigenstates of the density matrix $\rho$ describing the particle inside the interferometer right after the first transformation as $\ket{r}$, i.e., $\rho = \sum_r p_r \proj{r}$.
We also write $\mathcal{U}_2 \ket{r} = \ket{\phi_r}$.
From the definition of the probability of an inconclusive result,
\begin{equation}
\frac{P_?}{1-P_*} = \Tr \left( \sum_s \proj{s} \mathcal{U}_2 \, \tilde \rho \, \mathcal{U}_2^\dagger \right),
\label{APP_EQ_INC}
\end{equation}
where the sum is over the paths $\ket{s}$ at the output of the interferometer where the particle could be found if there was no bomb, i.e., if $\mathcal{U}_2 \, \rho \, \mathcal{U}_2^\dagger = \sum_r p_r \proj{\phi_r}$ is the state at the output.
Therefore, states $\ket{s}$ span a subspace that contains the eigenstates $\ket{\phi_r}$ and we conclude that,
\begin{equation}
\sum_s \proj{s} = \sum_r \proj{\phi_r} + \sum_\mu \proj{\mu},
\end{equation}
where the $\ket{\mu}$'s complement the subspace spanned by the paths.
Since $\bra{\mu} \mathcal{U}_2 \, \tilde \rho \, \mathcal{U}_2^\dagger \ket{\mu} \ge 0$, Eq.~(\ref{APP_EQ_INC}) admits the lower bound:
\begin{equation}
\frac{P_?}{1-P_*} \ge \Tr \left( \sum_r \proj{\phi_r} \mathcal{U}_2 \, \tilde \rho \, \mathcal{U}_2^\dagger \right) = \Tr \left( \sum_r \proj{r} \tilde \rho \right).
\end{equation}
Using the definition of $\tilde \rho$ in terms of $\rho$ given in Eq.~(\ref{EQ_TILDER}) of the main text we obtain
\begin{equation}
P_? \ge 1 - 2 P_* + P_* \bra{1} E(\rho) \ket{1},
\end{equation}
with $E(\rho) = \sum_r \proj{r}$.

\section{Proof of the cubes trade-off}
\label{SEC_APP_CUBES}

Let us first recall our assumption about cube $C$ describing the particle inside the interferometer
[Eq.~(\ref{EQ_CUBESASSUMPTION}) of the main text]:
\begin{equation}
\mathcal{T}_2(C) = \sum_s p_s M_s.
\label{APP_EQ_ASSUMEC}
\end{equation}
The following steps form the first part of the derivation:
\begin{align}
\frac{P_?}{1-P_*} & = \sum_s (M_s, \mathcal{T}_2 (\tilde C)) \nonumber \\
& \ge  \sum_s ( p_s M_s, \mathcal{T}_2 (\tilde C))\nonumber\\
& = ( \mathcal{T}_2 (C), \mathcal{T}_2 (\tilde C)) = (C, \tilde C).
\label{APP_EQ_FIRSTPART}
\end{align}
The first line is the definition of the probability of an inconclusive result,
the inequality follows from convexity, and then we used (\ref{APP_EQ_ASSUMEC}) and finally the fact that $\mathcal{T}_2$ preserves the inner product.
In the second part we shall find the minimum of the right-hand side.
Using the expression for the elements of $\tilde C$ in terms of the elements of $C$ we find:
\begin{equation}
(C, \tilde C) = \frac{1}{1-P_*} \sum_{j,k,l = 2}^N |C_{jkl}|^2,
\end{equation}
where $C_{222} + \dots + C_{NNN} = 1 - P_*$.
Since all of the summands are non-negative, we get the lower bound by setting all the off-diagonal terms to zero.
It is then easy to verify that the minimum is achieved for an even distribution of the probability:
\begin{equation}
C_{nnn}  = \frac{1-P_*}{N-1} \quad \textrm{ for } n = 2,\dots, N. 
\end{equation}
Using this lower bound in (\ref{APP_EQ_FIRSTPART}) we obtain
\begin{equation}
P_? \ge \frac{(1-P_*)^2}{N-1}.
\end{equation}

\section{Solution to the orthonormality equations for optimal cubes}
\label{SEC_APP_ONEQ}

The solution is divided into two parts: $N$ even and $N$ odd.

\subsection{$N$ even}

Let $M$ be a $(N-1) \times N$ matrix formed from the discrete $N \times N$ Fourier transform by deleting the first row:
\begin{equation}
M = \left(
\begin{array}{ccccc}
1&\omega^{1\cdot1}&\omega^{1\cdot2}&\cdots&\omega^{1\cdot (N-1)}\\
1&\omega^{2\cdot1}&\omega^{2\cdot2}&\cdots&\omega^{2\cdot (N-1)}\\
\vdots&\vdots&\vdots&\ddots&\vdots\\
1&\omega^{(N-1)\cdot1}&\omega^{(N-1)\cdot2}&\cdots&\omega^{(N-1)\cdot (N-1)}
\end{array}
\right),
\end{equation}
where $\omega = \exp (i 2\pi /N)$.
The crucial property we shall use is expressed in the following multiplication:
\begin{equation}
M^\dagger M =
\left(
\begin{array}{cccc}
N-1&-1&\cdots&-1\\
-1&N-1&\cdots&-1\\
\vdots&\vdots&\ddots&\vdots\\
-1&-1&\cdots&N-1\\
\end{array}
\right).
\end{equation}
Hence, the columns of matrix $M$ form vectors with fixed overlap equal to $-1$, for any pair of distinct vectors.
Let us now form the matrix $X$ by stacking $(N-2)/2$ matrices $M$ vertically:
\begin{equation}
X = \left( \begin{array}{c}
M \\
M \\
\vdots \\
M
\end{array}
\right).
\end{equation}
Note that matrix $X$ has $N$ columns and $(N-1)(N-2)/2$ rows.
We therefore define vectors $\vec x_n$ as columns of $X$:
\begin{equation}
X =\left(
\begin{array}{cccc}
\vec{x}_1&\vec{x}_2&\cdots&\vec{x}_N
\end{array}
\right).
\end{equation}
Indeed, every component of each $\vec x_n$ has unit modulus and appropriate overlap:
\begin{eqnarray}
& &(\vec{x}_{m},\vec{x}_{n}) + (\vec{x}_{n},\vec{x}_{m}) = 2 (\vec{x}_{m},\vec{x}_{n}) \nonumber \\
& & = 2 (m\textrm{th row of } X^\dagger) (n\textrm{th row of } X) \nonumber \\
& & = (N-2) (m\textrm{th row of } M^\dagger) (n\textrm{th row of } M) \nonumber \\
& & = 2 - N.
\end{eqnarray}

\subsection{$N$ odd}
We now construct the matrix $M$, having dimensions $\frac{N-1}{2} \times N$, by deleting the first row of the $N \times N$ Fourier transform matrix and taking only the top $\frac{N-1}{2}$ rows left:
\begin{equation}
M = \left(
\begin{array}{ccccc}
1&\omega^{1\cdot1}&\omega^{1\cdot2}&\cdots&\omega^{1\cdot N}\\
1&\omega^{1\cdot1}&\omega^{1\cdot2}&\cdots&\omega^{1\cdot N}\\
\vdots&\vdots&\vdots&\ddots&\vdots\\
1&\omega^{\frac{N-1}{2}\cdot1}&\omega^{\frac{N-1}{2}\cdot2}&\cdots&\omega^{\frac{N-1}{2}\cdot N}
\end{array}
\right).
\end{equation}
This time we have:
\begin{equation}
M^\dagger M = \left(
\begin{array}{cccc}
\frac{N-1}{2}&-\frac{1}{2}&\cdots&-\frac{1}{2}\\
-\frac{1}{2}&\frac{N-1}{2}&\cdots&-\frac{1}{2}\\
\vdots&\vdots&\ddots&\vdots\\
-\frac{1}{2}&-\frac{1}{2}&\cdots&\frac{N-1}{2}\\
\end{array}
\right)
+ i \, (\textrm{imaginary part}).
\end{equation}
We form the matrix $X$ by stacking $N-2$ matrices $M$ vertically and define vectors $\vec x_n$ as columns of $X$, as before.
Indeed, the overlap between distinct vectors reads:
\begin{equation}
\begin{aligned}
& (\vec{x}_{m},\vec{x}_{n}) + (\vec{x}_{n},\vec{x}_{m}) = 2 \mathrm{Re}\left[ (\vec{x}_{m},\vec{x}_{n})\right] \\
& = 2 \mathrm{Re} \left[ (m\textrm{th row of } X^\dagger) (n\textrm{th row of } X) \right] \\
&= 2 (N-2) \mathrm{Re}\left[ (m\textrm{th row of } M^\dagger) (n\textrm{th row of } M) \right] \\
&= 2 - N.
\end{aligned}
\end{equation}

\onecolumngrid
\subsection{Example of resulting cubes for $N = 4$}

The following four cubes are obtained for the four-path interferometer:

\begin{eqnarray}
C^{(1)} & = & \frac{1}{3\sqrt{3}}\left\{
\left(
\begin{array}{cccc} 
0 & 0 & 0 & 0\\
0 & 0 & 1 & 1  \\
0 &1& 0 &1 \\
0 & 1 & 1 & 0 
\end{array} \right),
\left(\begin{array}{cccc}
0 & 0 & 1  & 1   \\
0 & \sqrt{3} & 0 & 0 \\
1  & 0 & 0 & 0 \\
1  & 0 & 0 & 0
\end{array} \right),
\left(\begin{array}{cccc}
0 & 1  & 0 & 1   \\
1  & 0 & 0 & 0 \\
0 & 0 & \sqrt{3} & 0 \\
1  & 0 & 0 & 0
\end{array} \right),
\left(\begin{array}{cccc}
0 & 1  & 1  & 0\\
1  & 0 & 0 & 0 \\
1  & 0 & 0 & 0\\
0 & 0 & 0 & \sqrt{3}
\end{array}\right)
\right\},
\\
C^{(2)} & = & \frac{1}{3\sqrt{3}}\left\{
\left(\begin{array}{cccc}
\sqrt{3} & 0 & 0 & 0\\
0 & 0 & -i & -1  \\
0 & i & 0 & i \\
0 & -1 & -i & 0 
\end{array} \right),
\left(\begin{array}{cccc}
0 & 0 & i & -1  \\
0 & 0 & 0 & 0 \\
-i & 0 & 0 & 0 \\
-1 & 0 & 0 & 0
\end{array}\right),
\left(\begin{array}{cccc}
0 & -i & 0 & i  \\
i & 0 & 0 & 0 \\
0 & 0 & \sqrt{3} & 0 \\
-i & 0 & 0 & 0
\end{array}\right),
\left(\begin{array}{cccc}
0 & -1 & -i & 0\\
-1 & 0 & 0 & 0 \\
i & 0 & 0 & 0\\
0 & 0 & 0 & \sqrt{3}
\end{array}\right)
\right\}, \nonumber 
\end{eqnarray}
\begin{eqnarray}
C^{(3)} & = & \frac{1}{3\sqrt{3}}\left\{
\left(
\begin{array}{cccc}
\sqrt{3} & 0 & 0 & 0\\
0 & 0 & -1 & 1  \\
0 & -1 & 0 & -1 \\
0 & 1 & -1 & 0 
\end{array}\right),
\left(\begin{array}{cccc}
0 & 0 & -1 & 1  \\
0 & \sqrt{3} & 0 & 0 \\
-1 & 0 & 0 & 0 \\
1 & 0 & 0 & 0
\end{array}\right),
\left(\begin{array}{cccc}
0 & -1 & 0 & -1  \\
-1 & 0 & 0 & 0 \\
0 & 0 & 0 & 0 \\
-1 & 0 & 0 & 0
\end{array}\right),
\left(\begin{array}{cccc}
0 & 1 & -1 & 0\\
1 & 0 & 0 & 0 \\
-1 & 0 & 0 & 0\\
0 & 0 & 0 & \sqrt{3}
\end{array}\right)
\right\}, \nonumber
\\
C^{(4)} & = & \frac{1}{3\sqrt{3}}\left\{
\left(\begin{array}{cccc}
\sqrt{3} & 0 & 0 & 0\\
0 & 0 & i & -1  \\
0 & -i & 0 & -i \\
0 & -1 & i & 0 
\end{array}\right),
\left(\begin{array}{cccc}
0 & 0 & -i & -1  \\
0 & \sqrt{3} & 0 & 0 \\
i & 0 & 0 & 0 \\
-1 & 0 & 0 & 0
\end{array}\right),
\left(\begin{array}{cccc}
0 & i & 0 & -i  \\
-i & 0 & 0 & 0 \\
0 & 0 & \sqrt{3} & 0 \\
i & 0 & 0 & 0
\end{array}\right),
\left(\begin{array}{cccc}
0 & -1 & i & 0\\
-1 & 0 & 0 & 0 \\
-i & 0 & 0 & 0\\
0 & 0 & 0 & 0
\end{array} \right)
\right\}. \nonumber
\end{eqnarray}

\twocolumngrid

\section{Consistency of the transformation}
\label{SEC_APP_T}

\subsection{Positivity of matrix $D$}

We shall find the eigenvalues of $\mathds{1} - BB^\dagger$ explicitly.
Since $B = (X \, \, X^*)^T$, the construction in Appendix~\ref{SEC_APP_ONEQ} for even $N$ produces matrix $(N-1)^2 BB^\dagger$, given by
\begin{equation}
\left(
\begin{array}{c|c}
\begin{array}{ccc}
MM^\dagger & \cdots & MM^\dagger\\
\vdots & \ddots & \vdots\\
MM^\dagger & \cdots & MM^\dagger
\end{array}
& \begin{array}{ccc}
M^*M^\dagger & \cdots & M^*M^\dagger\\
\vdots & \ddots & \vdots\\
M^*M^\dagger & \cdots & M^*M^\dagger
\end{array}\\
\hline
\begin{array}{ccc}
MM^T & \cdots & MM^T\\
\vdots & \ddots & \vdots\\
MM^T & \cdots & MM^T
\end{array} & 
\begin{array}{ccc}
M^*M^T & \cdots & M^*M^T\\
\vdots & \ddots & \vdots\\
M^*M^T & \cdots & M^*M^T
\end{array}
\end{array}
\right),
\end{equation}
Note that $MM^\dagger = N \, \mathds{1}$ is inherited from the unitarity of the Fourier matrix.
Similarly, $M^*M^T = (M^\dagger)^T M^T = (M M^\dagger)^T = N \, \mathds{1}$.
Direct computation shows that $MM^T=M^*M^\dagger = N \, \mathds{1}^A$, where $\mathds{1}^A$ denotes an antidiagonal matrix with all nonzero elements being $1$.
Therefore, the matrix $(N-1)^2 B B^\dagger$ has the following form for even $N$:
\begin{equation}
\left(
\begin{array}{c|c}
\begin{array}{ccccccc}
1 & & &  & 1 & & \\
 & \ddots & & \cdots & & \ddots & \\
 & & 1 & & & & 1 \\
& \vdots & & & & \vdots & \\
1 & & &  & 1 & & \\
 & \ddots & & \cdots & & \ddots & \\
 & & 1 & & & & 1 \\ 
\end{array}
& 
\begin{array}{ccccccc}
 & & 1 &  &  & & 1 \\
 & \reflectbox{$\ddots$} & & \cdots & & \reflectbox{$\ddots$} & \\
1 & & & & 1 & & \\
& \vdots & & & & \vdots & \\ 
 & & 1 &  &  & & 1 \\
 & \reflectbox{$\ddots$} & & \cdots & & \reflectbox{$\ddots$} & \\
1 & & & & 1 & & \\ 
\end{array}
\\
\hline
\begin{array}{ccccccc}
 & & 1 &  &  & & 1 \\
 & \reflectbox{$\ddots$} & & \cdots & & \reflectbox{$\ddots$} & \\
1 & & & & 1 & & \\
& \vdots & & & & \vdots & \\
 & & 1 &  &  & & 1 \\
 & \reflectbox{$\ddots$} & & \cdots & & \reflectbox{$\ddots$} & \\
1 & & & & 1 & & \\ 
\end{array}
&
\begin{array}{ccccccc}
1 & & &  & 1 & & \\
 & \ddots & & \cdots & & \ddots & \\
 & & 1 & & & & 1 \\ 
& \vdots & & & & \vdots & \\
1 & & &  & 1 & & \\ 
 & \ddots & & \cdots & & \ddots & \\
 & & 1 & & & & 1 \\ 
\end{array}
\end{array}
\right).
\end{equation}
Similarly, one finds the following for odd $N$:
\begin{equation}
\left(
\begin{array}{c|c}
\begin{array}{ccccccc}
1 & & &  & 1 & & \\
& \ddots & & \cdots & & \ddots & \\
& & 1 & & & & 1 \\
& \vdots & & & & \vdots & \\
1 & & &  & 1 & & \\
& \ddots & & \cdots & & \ddots & \\
& & 1 & & & & 1 \\ 
\end{array}
& 
\begin{array}{c}
0
\end{array}
\\
\hline
\begin{array}{c}
0
\end{array}
&
\begin{array}{ccccccc}
1 & & &  & 1 & & \\
& \ddots & & \cdots & & \ddots & \\
& & 1 & & & & 1 \\
& \vdots & & & & \vdots & \\
1 & & &  & 1 & & \\
& \ddots & & \cdots & & \ddots & \\
& & 1 & & & & 1 \\ 
\end{array}
\end{array}
\right).
\end{equation}
In both cases, the eigenvalues of $BB^\dagger$ are either $0$ or $N(N-2)/(N-1)^2$.
Hence, the eigenvalues of $D$ are either $1$ or $1/(N-1)^2$.

\subsection{Exemplary transformation for $N=4$}
We shall only present the $D$ matrices. Following the method above one finds
\begin{equation}
	D = \frac{1}{3}
	\begin{pmatrix}
	2 & 0 & 0 & 0 & 0 & -1\\
	0 & 2 & 0 & 0 & -1 & 0\\
	0 & 0 & 2 & -1 & 0 & 0\\
	0 & 0 & -1 & 2 & 0 & 0\\
	0 & -1 & 0 & 0 & 2 & 0\\
	-1 & 0 & 0 & 0 & 0 & 2
	\end{pmatrix}.
\end{equation}
This is not the only solution given the constraints $\mathcal{T}^2 = \mathds{1}$ and $\mathcal{T} = \mathcal{T}^\dagger$.
The following two matrices were obtained by other means:
\begin{equation}
	D_2 = \frac{1}{3} \left(
	\begin{array}{cccccc}
	1 & -i & i & -i & i & 0\\
	i & 1 & 1 & -1 & 0 & -i\\
	-i & 1 & 1 & 0 & -1 & i\\
	i & -1 & 0 & 1 & 1 & -i\\
	-i & 0 & -1 & 1 & 1 & i\\
	0 & i & -i & i & -i & 1  
	\end{array} \right),
\end{equation}
\begin{equation}
	D_3 = \frac{1}{3} \left(
	\begin{array}{cccccc}
	1 & -1 & -i & i & 1 & 0\\
	-1 & 1 & -i & i & 0 & 1\\
	i & i & 1 & 0 & -i & -i\\
	-i & -i & 0 & 1 & i & i\\
	1 & 0 & i & -i & 1 & -1\\
	0 & 1 & i & -i & -1 & 1  
	\end{array} \right).
\end{equation}
If we use $D_3$ as an example, then $\mathcal{T}$ has the following elements:
\begin{equation}
	\mathcal{T} = \frac{1}{3}
	\begin{pmatrix}
	 0 & 1 & 1 & 1 & 1 & 1 & 1 & 1 & 1 & 1\\ 
	 1 & 0 & 1 & 1 & i & -1 & -i & -i & -1 & i \\ 
	 1 & 1 & 0 & 1 & -1 & 1 & -1 & -1 & 1 & -1 \\ 
	 1 & 1 & 1 & 0 & -i & -1 & i & i & -1 & -i \\
	 1 & -i & -1 & i & 1 & -1 & -i & i & 1 & 0 \\ 
	 1 & -1 & 1 & -1 & -1 & 1 & -i & i & 0 & 1 \\ 
	 1 & i & -1 & -i & i & i & 1 & 0 & -i & -i \\ 
	 1 & i & -1 & -i & -i & -i & 0 & 1 & i & i \\ 
	 1 & -1 & 1 & -1 & 1 & 0 & i & -i & 1 & -1\\ 
	 1 & -i & -1 & i & 0 & 1 & i & -i & -1 & 1 
	 \end{pmatrix}.
\end{equation}
\vspace{4mm}


\end{document}